\documentclass{aastex}

\begin{document}

\title{All solutions for geodesic anisotropic spherical collapse with shear and
heat radiation}
\author{B.V.Ivanov}
\affil{Institute for Nuclear Research and Nuclear Energy, \\
Bulgarian Academy of Science, \\
Tzarigradsko Shausse 72, Sofia 1784, Bulgaria}

\begin{abstract}
We introduce a physically important object, called the horizon function, in
the study of geodesic collapse. It is closely related to the stellar
characteristics and satisfies a simple Riccati equation. This equation is
integrated and all of its solutions are found in terms of some generating
function. Previous solutions are regained and further investigated.
\end{abstract}
\maketitle

\section{Introduction}

Gravitational collapse is an important issue in relativistic astrophysics.
There are many indications that the collapsing fluid in the star models is
anisotropic \cite{her 97}. In addition, this process is highly dissipative,
required to account for the enormous binding energy of the resulting object 
\cite{her 06}. Thus a realistic scenario is the collapse with heat flow \cite
{bon 89} or pure radiation \cite{tew 10}. Spherical collapse is described in
the general case by a diagonal metric with three independent components. For
simplicity, shearless fluid is used quite often, which reduces the metric
components to two. Even in the isotropic case the amount of interior
solutions is enormous \cite{iva 12}. The exterior solution is the Vaidya
shining star \cite{vai 51}. The main junction condition states that at the
star surface the radial pressure should equal the heat flux. This gives a
non-linear differential equation in partial derivatives (along radius and
time) and follows from the matching of the second fundamental forms. In the
streaming approximation (pure radiation) it leads to the vanishing of the
radial pressure.

In the shearless case the differential equation involves only the two metric
components $g_{00}$ and $g_{rr}$. When shear is present, the general three
component metric should be used, which further complicates the differential
equation. One can reduce in a different way the metric components to two, $%
g_{rr}$ and $g_{\theta \theta }$, by studying the geodesic case, $g_{00}=1$ 
\cite{kol 88}.

Interior anisotropic geodesic solutions with shear and without radiation
have been discussed in \cite{iva 11}. No matching to the exterior
Schwarzschild solution was done. The same problem in non-comoving
coordinates, but with matching was solved in \cite{her 02}

The first exact solution with radiation was obtained by \cite{nai 06}. After
that \cite{raj 08} noticed that the junction condition is a Riccati equation
for $g_{rr}$. They found two simple regular solutions in separated
variables. The solution of \cite{nai 06} is regained when certain parameters
are set to zero.

Later, \cite{thi 10} found even more general exact solutions depending on
arbitrary functions of the coordinate radius. They encompass the previous
solutions.

In a recent development \cite{abe 14} further expanded the realm of analytic
solutions by studying the Lie point symmetries of the boundary condition.
Generalized travelling waves and self-similar solutions were found.

The junction condition has been investigated also in the non-geodesic case.
For example a number of conformally flat solutions have been found \cite{her
04}, \cite{mah 05}, \cite{her 06}

The main idea of the present paper is to find and integrate an equation for
a physically meaningful object, which we call the horizon function. It is
directly related to the redshift and the formation of a horizon, which means
the appearance of a black hole as the end product of collapse. It enters the
expression for the mass of the star, the heat flow and the luminosity at
infinity. The equation is so simple that it is easily integrated with the
help of a generating function.

In Sect. 2 we present the Einstein equations, which in the anisotropic case
are expressions for the energy density, the radial and the tangential
pressure and the heat flow. The definition of the shear, the expansion, the
horizon function and the redshift are given. The relation between the mass
of the star and the horizon function is clarified. The main results of the
matching to the exterior Vaidya solution are shown. The most important of
them is a differential equation, involving the metric components. We show
that it is the same in the diffusion and in the streaming approximations.
The other stellar characteristics are also given. In Sect. 3, modifying the
method of \cite{thi 10}, the junction equation is written as a Riccati
equation for the horizon function with simple coefficients. This allows to
derive all geodesic solutions from two generating functions. Simple
expressions are given for the mass and its time derivative. It is shown how
the horizon function and even the redshift may be taken as alternative
generating functions. In Sect. 4 previously found solutions with radiation
are derived from the general solution and further investigated. In Sect. 5
we regain the solutions without radiation. Sect. 6 contains conclusions.

\section{Stellar characteristics}

The collapse of an anisotropic geodesic fluid sphere with shear is described
by the following metric 
\begin{equation}
ds^2=-dt^2+B^2dr^2+R^2\left( d\theta ^2+\sin ^2\theta d\varphi ^2\right) ,
\label{one}
\end{equation}
where $B$ and $R$ are independent functions of time $t$ and the radius $r$.
The spherical coordinates are numbered as $x^0=t$, $x^1=r$, $x^2=\theta $
and $x^3=\varphi $. The energy-momentum tensor, describing dissipation
through heat flow and null fluid, reads 
\begin{equation}
T_{ik}=\left( \mu +p_t\right) u_iu_k+p_tg_{ik}+\left( p_r-p_t\right) \chi
_i\chi _k+q_iu_k+u_iq_k+\varepsilon l_il_k.  \label{two}
\end{equation}
Here $\mu $ is the energy density, $p_r$ is the radial pressure, $p_t$ is
the tangential pressure, $u^i$ is the four-velocity of the fluid, $\chi ^i$
is a unit spacelike vector along the radial direction, $q^i$ is the heat
flow vector, also in the radial direction, $\varepsilon $ is the energy
density of the radiated null fluid and the vector $l^i$ is null. In comoving
coordinates we have 
\begin{equation}
u^i=\delta _0^i,\quad \chi ^i=B^{-1}\delta _1^i,\quad q^i=q\chi ^i,\quad
l^i=u^i+\chi ^i.  \label{three}
\end{equation}
The Einstein field equations become \cite{thi 10}, \cite{iva 10} 
\begin{equation}
\mu +\varepsilon =\left( \frac{2\dot B}B+\frac{\dot R}R\right) \frac{\dot R}%
R-\frac 1{B^2}\left( \frac{2R^{\prime \prime }}R+\frac{R^{\prime 2}}{R^2}-%
\frac{2B^{\prime }R^{\prime }}{BR}-\frac{B^2}{R^2}\right) ,  \label{four}
\end{equation}
\begin{equation}
p_r+\varepsilon =-\frac{2\ddot R}R-\frac{\dot R^2}{R^2}+\frac{R^{\prime 2}}{%
B^2R^2}-\frac 1{R^2},  \label{five}
\end{equation}
\begin{equation}
p_t=-\left( \frac{\ddot B}B+\frac{\ddot R}R+\frac{\dot B\dot R}{BR}\right)
+\frac 1{B^2}\left( \frac{R^{\prime \prime }}R-\frac{B^{\prime }R^{\prime }}{%
BR}\right) ,  \label{six}
\end{equation}
\begin{equation}
qB+\varepsilon =-\frac 2B\left( \frac{\dot BR^{\prime }}{BR}-\frac{\dot
R^{\prime }}R\right) .  \label{seven}
\end{equation}
Here the dot means a time derivative, while the prime stands for a radial
derivative.

For the line element (1) the four-acceleration vanishes, while the shear and
the expansion scalars are given by 
\begin{equation}
\sigma =\frac 13\left( \frac{\dot R}R-\frac{\dot B}B\right) ,  \label{eight}
\end{equation}
\begin{equation}
\Theta =\frac{2\dot R}R+\frac{\dot B}B.  \label{nine}
\end{equation}
Next, we introduce the important object $H$, which we call ''the horizon
function'' for reasons to become clear later: 
\begin{equation}
H=\frac{R^{\prime }}B+\dot R.  \label{ten}
\end{equation}
The mass entrapped within radius $r$ is given by the expression \cite{cah 70}
\begin{equation}
m=\frac R2\left[ 1+\dot R^2-\left( \frac{R^{\prime }}B\right) ^2\right] .
\label{eleven}
\end{equation}
On the stellar surface $\Sigma $ it becomes the mass of the star. The
compactness parameter reads $u=m/R$. Eq (11) can be rewritten using $H$%
\begin{equation}
\frac{2m}R=1+2H\dot R-H^2.  \label{twelve}
\end{equation}

The exterior spacetime is given by the Vaidya shining star solution 
\begin{equation}
ds^2=-\left[ 1-\frac{2M\left( v\right) }\rho \right] dv^2-2dvd\rho +\rho
^2\left( d\theta ^2+\sin ^2\theta d\varphi ^2\right) ,  \label{thirt}
\end{equation}
where $M\left( v\right) $ is the mass of the star measured at time $v$ by an
observer at infinity, while $\rho $ is the exterior coordinate radius. Both
solutions should be joined smoothly at $\Sigma $, which leads to the
following junction conditions: 
\begin{equation}
R_\Sigma =\rho _\Sigma \left( v\right) ,  \label{fourt}
\end{equation}
\begin{equation}
m_\Sigma =M_\Sigma ,  \label{fift}
\end{equation}
\begin{equation}
\left( p_r\right) _\Sigma =\left( qB\right) _\Sigma .  \label{sixt}
\end{equation}
Eq (16) should be satisfied by $R$ and $B$ while the other equations are
definitions of different stellar characteristics. When $q$ vanishes, the
radial pressure should vanish at the surface, but when $\varepsilon \neq 0$
Eqs.(5,7) show that in terms of the metric Eq (16) is restored. This
condition was used in many works with null fluid radiation \cite{tew 10}.
When $\varepsilon $ also vanishes we get collapse without radiation and the
exterior solution is the Schwarzschild vacuum solution. In the following we
set $\varepsilon =0$.

Some important stellar characteristics are also defined on the surface of
the star. These are the redshift $z_\Sigma $%
\begin{equation}
z_\Sigma =\frac 1{H_\Sigma }-1,  \label{sevent}
\end{equation}
the surface luminosity $\Lambda _\Sigma $ and the luminosity at infinity $%
\Lambda _\infty $%
\begin{equation}
\Lambda _\Sigma =\left( \frac 12qBR^2\right) _\Sigma ,  \label{eighteen}
\end{equation}
\begin{equation}
\Lambda _\infty =H_\Sigma ^2\Lambda _\Sigma .  \label{ninet}
\end{equation}
The temperature at the surface is given by 
\begin{equation}
T_\Sigma ^4=\frac{\left( qB\right) _\Sigma }{8\pi \delta },  \label{twenty}
\end{equation}
where $\delta $ is some constant.

It is seen that the star properties have simpler expressions when written in
terms of $H$. The redshift is positive during collapse. Then Eq (17) shows
that $0\leq H_\Sigma \leq 1$. When $H_\Sigma =0$ we obtain from Eq (12) and
the junction conditions 
\begin{equation}
\left( 1-\frac{2M\left( v\right) }\rho \right) _\Sigma =0.  \label{twone}
\end{equation}
This signals the appearance of a horizon and a black hole within it, which
is the typical end of gravitational collapse. This explains the name of $H$.
The redshift becomes infinite, while the luminosity at infinity drops to
zero. The point in time when collapse starts is taken usually as $-\infty $.
There $H_\Sigma $ should have some positive value less or equal to $1$. Thus
during the collapse the horizon function decreases to zero and $\dot
H_\Sigma \leq 0$. Other scenarios will be discussed in the following.

\section{Solution of the junction equation}

With the help of Eqs. (5,7) the main junction condition (16) becomes 
\begin{equation}
2R\ddot R+\dot R^2+1-\frac{R^{\prime 2}}{B^2}+\frac{2R\dot R^{\prime }}B-%
\frac{2\dot BRR^{\prime }}{B^2}=0.  \label{twtwo}
\end{equation}
It and the following equations hold on the surface. This equation coincides
with Eq (9) from \cite{thi 10} and determines the evolution of a radiating
geodesic and anisotropic star with shear. It is a highly nonlinear
differential equation in partial derivatives. It is simplified when one
introduces the function $Z$%
\begin{equation}
Z=\frac B{R^{\prime }}  \label{twthree}
\end{equation}
and becomes 
\begin{equation}
\dot Z=\frac 1{2R}\left( FZ^2-1\right) ,  \label{twfour}
\end{equation}
where 
\begin{equation}
F=2R\ddot R+\dot R^2+1.  \label{twfive}
\end{equation}
Eq (24) is separable and integrable as long as $F$ does not depend on time.
The latter condition leads to an equation for $R$, which can be solved and
special solutions for $R$ and $B$ may be obtained.

In the general case, let us express $Z$ in terms of the horizon function.
Eqs (10, 23) yield 
\begin{equation}
\frac 1Z=H-\dot R.  \label{twsix}
\end{equation}
Now we insert this expression in Eq (24) and obtain after some calculations
and cancellations

\begin{equation}
2R\dot H=H^2-2\dot RH-1.  \label{twseven}
\end{equation}
Eq (27) is much simpler than Eq (22). It involves only first time
derivatives and the physical quantities $R$ (the radius of the star as seen
from an exterior observer) and the horizon function $H$ which controls the
black hole formation, the redshift, the mass and the luminosity of the star.
It is a Riccati equation with respect to $H$ and a linear one with respect
to $R$. We can do even better by defining the non-negative function $D=RH$.
Then Eq (27) transforms into 
\begin{equation}
\dot D=\frac 1{2R^2}D^2-\frac 12.  \label{tweight}
\end{equation}
$D$ still satisfies a Riccati equation, but $R$ enters in an algebraic way.
Thus we get an expression for $R$ in terms of $D$%
\begin{equation}
R=\frac D{\sqrt{2\dot D+1}}.  \label{twnine}
\end{equation}
In addition to $D\geq 0,$ $D$ should satisfy $\dot D>-1/2$.

Eq (29) holds on $\Sigma $, that is, $r=r_\Sigma $, which is some constant.
We can take any reasonable $D\left( t\right) $ and promote the constants in
it to arbitrary functions of the radius. This situation is similar to the
solution of the isotropic equation for perfect fluid without shear, when the
metric can be written in isotropic coordinates (which contains only radial
derivatives) \cite{iva 12}.

One difference is that the junction equation holds for any fluid, either
perfect or imperfect, as long as there is a clear boundary between the
interior and the exterior solution.

Another difference is that we can add to the r.h.s. of Eq (29) the function $%
P=g\left( r\right) F\left( t,r\right) $ where $g$ and $F$ are arbitrary up
to ensuring that $R$ is positive. In addition $f\left( r_\Sigma \right) =0$
and the term $P$ does not show on the surface (zero constant there). Thus
the continuation of $R$ in the bulk of the star depends on a second
generating function $P$. For the moment we set $g=0$ and shall comment on it
at the end of the section. Hence, Eq (29) and the following equations hold
in the bulk too.

From Eq (29) and the definition of $D$ we obtain an expression for $H$%
\begin{equation}
H=\sqrt{2\dot D+1}.  \label{thirty}
\end{equation}
Then Eq (10) gives an expression for $B$ 
\begin{equation}
B=\frac{R^{\prime }}{H-\dot R},  \label{thone}
\end{equation}
where we should insert the expressions for $R$ and $H$ in terms of $D$. In
the process of collapse $R$ decreases, hence $\dot R<0$ and the denominator
in Eq (31) is positive. Then it follows that $R^{\prime }>0$. This is
exactly the condition for the absence of shell crossing singularities \cite
{mal 11}.

We have determined the metric components in terms of $D$ and have solved the
junction equation. The arbitrary function $D\left( t.r\right) $ plays the
role of a generating function. All stellar characteristics become functions
of it and its time derivatives.

Eqs (29,30) lead to the useful formulas 
\begin{equation}
\dot R=\frac{\left( 2\dot D+1\right) \dot D-D\ddot D}{\left( 2\dot
D+1\right) ^{3/2}},  \label{thtwo}
\end{equation}
\begin{equation}
H-\dot R=\frac{\left( 2\dot D+1\right) \left( \dot D+1\right) +D\ddot D}{%
\left( 2\dot D+1\right) ^{3/2}},  \label{ththree}
\end{equation}
\begin{equation}
\dot H=\frac{\ddot D}{\sqrt{2\dot D+1}},  \label{thfour}
\end{equation}
hence, $\ddot D<0$. A combination of Eqs (12) and (27) gives a simple
formula for the mass function 
\begin{equation}
m=-R^2\dot H.  \label{thfive}
\end{equation}
In terms of $D$ it becomes 
\begin{equation}
m=-\frac{D^2\ddot D}{\left( 2\dot D+1\right) ^{3/2}}.  \label{thsix}
\end{equation}
Thus, while $H=0$ describes the formation of a black hole, $\dot H=0$
describes the burning out of the mass due to the radiation, which is another
reason for the end of collapse. The question is which one happens first. A
third scenario is the ''eternal collapse'' when $2m/R$ is a constant,
smaller than $1$ so that a horizon never develops. It is a well-known option
in the shearless case \cite{tew 15}.

Let us take next Eq (7), the expression for $qB$. It may be transformed into 
\begin{equation}
qB=-\frac{2R^{\prime }}{BR}\left( \ln \frac B{R^{\prime }}\right) ^{.}=\frac
2R\left( \frac{R^{\prime }}B\right) ^{.},  \label{thseven}
\end{equation}
which can be expressed through $H$%
\begin{equation}
qB=\frac 2R\left( H-\dot R\right) ^{.}.  \label{theight}
\end{equation}
The l.h.s. is positive because $q>0$ and the energy is radiated out.

On the other side Eq (35) yields 
\begin{equation}
\dot m=-R\left( 2\dot R\dot H+R\ddot H\right) .  \label{thnine}
\end{equation}
The time derivative of the junction equation, however, gives 
\begin{equation}
2\dot R\dot H+R\ddot H=H\left( H-\dot R\right) .^{.}  \label{forty}
\end{equation}
Comparing this to Eq (38) yields finally 
\begin{equation}
2\dot m=-qBR^2H.  \label{foone}
\end{equation}
This relation shows how the decrease in mass is governed by the heat flow.
In terms of the generating function $D$ we have 
\begin{equation}
qB=2\frac{\left( 2\dot D+1\right) \left( 2\dot D\ddot D+DD^{...}\right)
-3D\ddot D^2}{D\left( 2\dot D+1\right) ^2},  \label{fotwo}
\end{equation}
\begin{equation}
\dot m=\frac{D\left[ 3D\ddot D^2-\left( 2\dot D+1\right) \left( 2\dot D\ddot
D+DD^{...}\right) \right] }{\left( 2\dot D+1\right) ^{5/2}},  \label{fothree}
\end{equation}
Knowing $qB$ and $H$ we get expressions for the redshift, the two
luminosities and the surface temperature from Eqs (17-20). We have explained
that the junction equation holds for all $r$, so it gives 
\begin{equation}
p_r=qB  \label{fofour}
\end{equation}
and $p_r>0$ everywhere inside the star.

The other characteristics of the fluid will be given for simplicity in terms
of $R,$ $H$ and sometimes $B$. The shear and the expansion become 
\begin{equation}
\sigma =\frac 13\left[ \ln \frac R{R^{\prime }}\left( H-\dot R\right)
\right] ^{.},  \label{fofive}
\end{equation}
\begin{equation}
\Theta =\left( \ln \frac{R^2R^{\prime }}{H-\dot R}\right) ^{.}.
\label{fosix}
\end{equation}
The energy density $\mu $ is related to the mass 
\begin{equation}
\mu =\frac{2m}{R^3}+\frac 2{BR}\left[ \dot B\dot R-\left( H-\dot R\right)
^{\prime }\right] ,  \label{foseven}
\end{equation}
while the tangential pressure is 
\begin{equation}
p_t=\frac 12\left( \frac{2m}{R^3}-\mu \right) -\frac 1{BR}\left( \ddot
RB+R\ddot B\right) ,  \label{foeight}
\end{equation}

The function $D$ is not the only generation function. One can use $H$
instead. Eq (30) may be integrated to give 
\begin{equation}
2D=\int H^2dt-t.  \label{fonine}
\end{equation}
Inserting the definition $D=HR$ we obtain an expression for $R$%
\begin{equation}
R=\frac 1{2H}\left( \int H^2dt-t\right) .  \label{fifty}
\end{equation}
Taking the time derivative of this equation and using the junction equation
(27) we get 
\begin{equation}
\dot R=-\frac{\dot H}HR+\frac{H^2-1}{2H},  \label{fione}
\end{equation}
\begin{equation}
H-\dot R=\frac{\dot H}HR+\frac H2+\frac 1{2H}.  \label{fitwo}
\end{equation}
With the help of Eqs (50-52) we can reduce all expressions above in $R$ and $%
H$ to functions of $H$ and its derivatives, without going to the $D$ level.

The initial junction equation (16) is a relation between physical
characteristics of the model - the radial pressure and the heat flow. We can
do the same for the transformed equation (27), which already contains the
luminosity radius $R$. With the help of the Eq (17), where the constants are
promoted to functions of $r$ we can pass from $H$ to the redshift function $%
z\left( t,r\right) $ which on the surface is a physical characteristic of
the model. The result is once again a Riccati equation for $z$ and the
analogue of Eq (50) for $R$ 
\begin{equation}
\dot z=\frac 1{2R}z^2+\frac{1+\dot R}Rz+\frac{\dot R}R,  \label{fithree}
\end{equation}
\begin{equation}
R=2\left( 1+z\right) \left( \int \frac{dt}{\left( 1+z\right) ^2}-t\right) .
\label{fifour}
\end{equation}
One should use reasonable values for $z$ which are limited from above by the
energy conditions \cite{iva 02}.

What happens when $g\neq 0$? As long as a stellar characteristic depends on $%
D$ and on $R$ or/and its time derivatives, $P=0$ on the surface where many
of the important stellar characteristics are defined. It is seen that almost
all of the above expressions are of this type. The metric component $B$ is
an exception, but it has no physical meaning. Nevertheless, let us see how
the above formulas change in this case. We have 
\begin{equation}
R=R_0+P  \label{fifive}
\end{equation}
where $R_0$ is the star radius given by Eq (29), 
\begin{equation}
H=H_0\frac{R_0}{R_0+P}  \label{fisix}
\end{equation}
\begin{equation}
\frac 12qBR=H_0\frac{\dot R_0P-R_0\dot P}{\left( R_0+P\right) ^2}+\frac{\dot
H_0R_0}{R_0+P}-\ddot R_0-\ddot P  \label{fiseven}
\end{equation}
and so on.

\section{Previous solutions with heat radiation}

Now one can take different functions $D,$ $H$ or $z$ and try to find
realistic stellar models. It is interesting to obtain in the first place the
different particular solutions found in the past. The most general of them
were presented by \cite{abe 14}, using the Lie symmetry method. In this way
generalised travelling waves and self-similar solutions have been found,
which depend on an arbitrary function. In our approach we take $D=D\left(
x\right) $ where $x$ is 
\begin{equation}
x=\int \frac{dr}{f\left( r\right) }-\frac ta  \label{fieight}
\end{equation}
and $a$ is a constant, while $f\left( r\right) $ is an arbitrary function.
When $f\left( r\right) =1$ we have a travelling wave with speed $1/a$. One
easily finds that $\dot D=\dot D\left( x\right) $ and Eq (29) shows that $%
R=R\left( x\right) $. Also $H=H\left( x\right) $ and from Eq (31) 
\begin{equation}
B=\frac{R_x}{f\left( r\right) \left( H+R_x/a\right) }=\frac{h\left( x\right) 
}{f\left( r\right) }.  \label{finine}
\end{equation}
We obtain exactly the first class of solutions in \cite{abe 14}. Here $h$
satisfies a Riccati equation because $B$ does so, as seen from Eq (22).

Now let us take $D$ in the form 
\begin{equation}
D=D_1\left( x\right) t,\quad x=t\exp \left( -\int \frac{dr}{af\left(
r\right) }\right) .  \label{sixty}
\end{equation}
When $f=r/a$, $x=t/r$, that is, $x$ is a self-similar variable. Now $\dot
D=\dot D\left( x\right) $ and $R=g\left( x\right) t$, where $g$ is
arbitrary. Next $H=H\left( x\right) $ and 
\begin{equation}
B=\frac{-xR_x}{af\left( H-xg_x-g_x\right) }=\frac{h\left( x\right) }{f\left(
r\right) }\exp \int \frac{dr}{af\left( r\right) },  \label{sione}
\end{equation}
where $h$ satisfies a different Riccati equation. Thus we obtain the second
class of \cite{abe 14} solutions.

Next, let us go back to Eqs (24,25). Eq (25) may be written as 
\begin{equation}
F=\frac{\left( R\dot R^2\right) ^{.}}{\dot R}+1.  \label{sitwo}
\end{equation}
When $F=1+R_1\left( r\right) ^2$, with $R_1\left( r\right) $ arbitrary, the
solution of Eq (62) is 
\begin{equation}
R=R_1t+R_2,  \label{sithree}
\end{equation}
where $R_2\left( r\right) $ is another arbitrary function of $r$ \cite{thi
10}. This means that $\ddot R=0$. Since $H$ decreases, $\dot H\leq 0$ and,
e.g., the mass function (35) is positive. But then Eq (38) shows that $q\leq
0$, therefore the star absorbs radiation from outside, which is not
realistic. Consequently, the linear in time solution for $R$ is not physical.

When $F=1$ Eq (62) yields 
\begin{equation}
R=\left( R_1t+R_2\right) ^{2/3}.  \label{sifour}
\end{equation}
Now Eq (38) reads 
\begin{equation}
qBR=-\frac{2\dot Z}{Z^2}  \label{sifive}
\end{equation}
and thus $\dot Z<0$. Then Eq (24) shows that $Z^2<1$. It is easily seen \cite
{thi 10} that 
\begin{equation}
Z=\frac{1-y}{1+y},  \label{sisix}
\end{equation}
where 
\begin{equation}
y=-f\left( r\right) \exp \left[ 3\left( R_1t+R_2\right) ^{1/3}/R_1\right] .
\label{siseven}
\end{equation}
The above condition for $Z^2$ demands that $y$ is positive and, hence, $f$
is negative. Usually, collapse is supposed to take place for $-\infty <t\leq
0$ and at $t=0$ a black hole appears. Therefore we take on the surface $%
R_1<0 $ and $R_2>0$. Then $\dot R_\Sigma $ is really negative and the star
shrinks. As we have mentioned before, the condition for a black hole
formation is $H_{\Sigma 0}=0$. Eq (64) leads to 
\begin{equation}
\dot R=\frac 23\frac{R_1}{\sqrt{R}},\quad \ddot R=-\frac 29\frac{R_1^2}{R^2}.
\label{sieight}
\end{equation}

Then Eq (26) gives 
\begin{equation}
\frac{1+y_{\Sigma 0}}{1-y_{\Sigma 0}}=c,\quad c=\left( -\frac{2R_1}{%
3R_2^{1/3}}\right) _\Sigma >0,  \label{sinine}
\end{equation}
which is a linear equation for $y_{\Sigma 0}$ with solution 
\begin{equation}
y_{\Sigma 0}=-f\left( r_\Sigma \right) \exp \left( -\frac 2c\right) =\frac{%
c-1}{c+1}.  \label{seventy}
\end{equation}
This defines $f\left( r_\Sigma \right) $ as long as the constant $c>1$. At $%
t=-\infty $ we have $R=\infty $, $Z=1$, $H=1$, while at $t=0$ we find $%
R_0=R_2^{2/3},$ $Z_{\Sigma 0}=1/c$, $H_{\Sigma 0}=0$. In fact, the collapse
of the star may start at some finite negative time from a static model,
which develops for some reasons non-trivial heat flow.

Is it possible that the end state is not a black hole but Minkowski
spacetime, because all the mass has been burnt out and we get $m_\Sigma =0$?
This situation occurs in the non-geodesic collapse with shear when $R_\Sigma
=0$ \cite{pin 11}. Eq (35) shows that in the geodesic case this may happen
when $R_\Sigma =0$ or $\dot H_\Sigma =0$. Differentiating Eq (26) and using
Eq (24) we obtain 
\begin{equation}
\dot H=\frac 1{2R}\frac{1-Z^2}{Z^2}+\ddot R.  \label{seone}
\end{equation}
Inserting this in the mass formula and using Eq (68) yields 
\begin{equation}
m=R\frac{Z^2-1}{Z^2}+\frac 29R_1^2.  \label{setwo}
\end{equation}

When $t=-\left( R_2/R_1\right) _\Sigma $, which is positive, $R_\Sigma =0$.
Then Eqs (66, 67) show that $Z_\Sigma $ becomes a finite constant, while Eq
(68) gives $\dot R_\Sigma =-\infty $. Thus Eq (26) gives $H_\Sigma =-\infty $%
, well beyond its range of $[0,1]$. Eq (17) yields for the redshift $%
z_\Sigma =-1$, which is completely unrealistic. In addition, Eq (72) shows
that the mass does not vanish for $R_\Sigma =0$. Hence, the scenario of \cite
{pin 11} is not realized here. In fact, $H_\Sigma $, being a continuous
function in time, will pass through its zero before going to negative
infinity and there a black hole will be formed, stopping the collapse.

Let us explore the other possibility, namely, $\dot H_\Sigma =0$ when $t=0$.
When $\dot H=0$, Eq (71) gives 
\begin{equation}
Z^2\left( 1-2R\ddot R\right) =1.  \label{sethree}
\end{equation}
Going to the star surface and setting $t=0$ we obtain 
\begin{equation}
y_{\Sigma 0}=-f\left( r_\Sigma \right) e^{-\frac 2c}=\frac{\sqrt{1+c^2}-1}{%
\sqrt{1+c^2}+1},  \label{sefour}
\end{equation}
which is the analogue of Eq (70) and fixes $f\left( r_\Sigma \right) $. Eq
(26) gives 
\begin{equation}
H_{\Sigma 0}=\sqrt{1+c^2}-c\in (0,1].  \label{sefive}
\end{equation}
Hence, the star burns out ($m_{\Sigma 0}=0$) before the formation of horizon
and a black hole ($H_{\Sigma 0}>0$) and the process of collapse stops. The
interior solution becomes Minkowski spacetime with radius $\left(
R_2^{2/3}\right) _\Sigma $ and it is joined smoothly to the exterior
solution, which is again Minkowski spacetime ( $M=0$).

Eq (26) gives general expressions for the two generating functions $H$ and $%
D $: 
\begin{equation}
H=\frac 1Z+\dot R  \label{sesix}
\end{equation}
\begin{equation}
D=\frac RZ+R\dot R  \label{seseven}
\end{equation}

When $F=1$, $H$ becomes due to Eqs. (64, 66) 
\begin{equation}
H=\frac{1+y}{1-y}+\frac 23R_1\left( R_1t+R_2\right) ^{-1/3}  \label{seeight}
\end{equation}
where $y$ is given by Eq. (67). A similar expression holds for $D$.

When $F=1+R_1^2$, $H$ reads 
\begin{equation}
H=\sqrt{R_1^2+1}\frac{1+l}{1-l}+R_1  \label{senine}
\end{equation}
\begin{equation}
l=-f\left( r\right) \left( R_1t+R_2\right) ^{\sqrt{R_1^2+1}/R_1}
\label{eighty}
\end{equation}
where the formula for $l$ has been taken from \cite{thi 10}. Similar formula
holds for $D$.

Finally, there is a solution with vanishing luminosity $\Lambda _\Sigma $,
even though the heat flux does not vanish in the bulk (but vanishes on $%
\Sigma $ so that Eq (18) holds). It is called the generalized LTB solution 
\cite{her 10b}. In it $q$ is invoked wholly by the term $P$. As seen from Eq
(57) $q\sim g\left( r\right) $ when $q_0$ (given by Eq (38)) is zero and
vanishes on the surface, but not in the bulk of the star.

\section{Previous solutions without heat radiation}

Now let us make the passage to the case with $q=0$. Then Eqs (37, 65, 76)
yield 
\begin{equation}
\dot Z=0,\quad B=R^{\prime }/a\left( r\right) ,\quad H=a+\dot R
\label{eione}
\end{equation}
where $a\left( r\right) $ is some positive function of integration. Plugging
this form of $H$ into its Riccati equation (27) we get an equation for $R$%
\begin{equation}
2R\ddot R+\dot R^2+1=a^2  \label{eitwo}
\end{equation}
A look at Eq(5) (with $\varepsilon =0$) shows that this gives the vanishing
of the radial pressure $p_r=0$. This also follows from the main form of the
junction condition, Eq (16) when $q=0$. It naturally gives $F=a^2$, which
also follows from Eq (24) when $\dot Z=0$. Using the expression for $F$ from
Eq (59) we transform Eq (82) into 
\begin{equation}
\left( R\dot R^2\right) ^{.}=\left( a^2-1\right) \dot R  \label{eithree}
\end{equation}
Integration of this equation leads to 
\begin{equation}
\dot R^2=\frac{2m}R+k  \label{eifour}
\end{equation}
Here $m\left( r\right) $ is another function of integration and we have
introduced $k=a^2-1$. The meaning of $m$ is seen from Eq (12). When $q=0$,
Eq (41) shows that the mass is a function of $r$ only. Plugging $H$ from Eq
(81) into Eq (12) results into Eq (84) with $m$ being the mass function of
the star.

Interior anisotropic geodesic solutions with shear and without radiation
have been discussed in \cite{iva 11}. Eq (82) follows from Eq (19) in this
reference when $p_r=0$. The same problem in non-comoving coordinates, but
with matching was solved in \cite{her 02}. Eq (81) for $B$ and Eq (84)
characterize the LTB solution \cite{her 10b}, which has three cases, but
here it arises from an anisotropic source. When the fluid is perfect ($%
p_r=p_t$), both pressures vanish and the general dust solution is obtained.

Let us consider now the so called Euclidean star solutions \cite{her 10a}.
They satisfy the constraint $B=R^{\prime }.$ Then Eq (10) becomes 
\begin{equation}
H-\dot R=1.  \label{eifive}
\end{equation}
Now Eq (28) yields $q=0$, hence, the geodesic Euclidean stars are
non-radiative. This also follows from Eq (39) in the previous reference.
They fall in the parabolic LTB class because $a=1$ and hence $k=0$.

The generating function of the LTB solutions is rather simple 
\begin{equation}
D=R\left( \sqrt{1+k}+\dot R\right)  \label{eisix}
\end{equation}
where $R$ satisfies Eq (84).

\section{Conclusions}

In this paper we have integrated the main junction condition (16, 22) for
geodesic anisotropic spherical collapse and expressed all possible solutions
through the generating functions $D=RH$ and $P$. We have also shown that $H$
or the redshift may play the role of alternative generating functions. For
this purpose we have introduced the physically important object $H$, called
the horizon function. Fortunately, it satisfies a Riccati equation, simpler
than the previous such equations for $B$ and $Z$. It rules the appearance of
a black hole ($H=0$) or the vanishing of the star's mass due to radiation ($%
\dot H=0$).

All previous exact solutions have been regained. Those obtained by the Lie
symmetry method \cite{abe 14} are based on $D$ depending on some function of 
$t,r$.

The solution with linear in time physical radius $R$ leads in the shearless
case to infinite collapse without black hole \cite{ban 02}, \cite{iva 12}, 
\cite{tew 15}. We have shown that when shear is present it has negative heat
flow and is physically unrealistic.

Another class of solutions of \cite{thi 10} appears to be realistic.
Depending on the integration constants their collapse may lead to a black
hole or Minkowski spacetime due to the burning away of the star's mass.

We have also regained the LTB solutions, the generalised LTB solutions \cite
{her 10b} and the geodesic Euclidean stars \cite{her 10a}, which must be
non-radiative.

Certainly, the use of generating functions will lead in the future to the
discovery of other realistic star models, describing anisotropic geodesic
collapse with radiation.

\end{document}